\documentclass[aip,reprint,graphicx]{revtex4-1}

\usepackage[american]{babel}

\usepackage{grffile}

\usepackage{newtxtext,newtxmath}
\usepackage{microtype}

\usepackage{amsmath}
\usepackage{amssymb}
\usepackage{bbm}
\usepackage{mathtools}
\usepackage{dsfont}
\usepackage{braket}
\usepackage{cancel}
\usepackage{slashed}

\usepackage{graphicx}
\usepackage[svgnames]{xcolor}
\usepackage{transparent}

\usepackage{siunitx}

\usepackage{ragged2e}
\usepackage{array}
\usepackage{tabularx}
\usepackage{booktabs}
\usepackage{makecell}

\definecolor{cset-aps-blueberry}{RGB}{28,128,158}
\definecolor{cset-aps-blue}{RGB}{46,44,184}
\definecolor{cset-aps-turquoise}{RGB}{0,67,88}
\definecolor{cset-aps-limegreen}{RGB}{190,219,67}
\definecolor{cset-aps-green}{RGB}{31,138,112}
\definecolor{cset-aps-yellow}{RGB}{255,225,25}
\definecolor{cset-aps-orange}{RGB}{253,116,0}
\definecolor{cset-aps-red}{RGB}{219,0,43}

\usepackage{tikz}

\usepackage{pgfplots}
\pgfplotsset{%
	every axis legend/.append style={%
		cells={anchor=west},
		at={(0.96,0.04)},
		anchor=south east,
		font=\scriptsize,
	},
	every axis/.append style={%
		yticklabel style={%
			/pgf/number format/fixed zerofill,
			/pgf/number format/precision=2},
	},
	width= \textwidth,
	height=8cm,
	xmajorgrids=true,
	xminorgrids=false,
	minor x tick num=1,
}
\usepgfplotslibrary{external}

\usetikzlibrary{decorations}

\usepackage{pict2e,picture}

\makeatletter
\DeclareRobustCommand{\Arrow}[1][]{%
	\check@mathfonts
	\if\relax\detokenize{#1}\relax
	\settowidth{\dimen@}{$\m@th\rightarrow$}%
	\else
	\setlength{\dimen@}{#1}%
	\fi
	\sbox\z@{\usefont{U}{lasy}{m}{n}\symbol{41}}%
	\begin{picture}(\dimen@,\ht\z@)
		\roundcap
		\put(\dimexpr\dimen@-.7\wd\z@,0){\usebox\z@}
		\put(0,\fontdimen22\textfont2){\line(1,0){\dimen@}}
	\end{picture}%
}
\makeatother

\usepackage{hyperref}
\hypersetup{%
	colorlinks=true,
	linkcolor={cset-aps-red},
	linkbordercolor={cset-aps-red},
	filecolor={cset-aps-orange},
	filebordercolor={cset-aps-orange},
	citecolor={cset-aps-blue},
	citebordercolor={cset-aps-blue},
	urlcolor={cset-aps-green},
	urlbordercolor={cset-aps-green},
	menucolor={cset-aps-limegreen},
	menubordercolor={cset-aps-limegreen},
	breaklinks=true,
	pdfborderstyle={/S/U/W 2},
	pdfpagemode=UseOutlines,
	pdfstartpage={1},
}

\newcommand{\ee}{e}
\newcommand{\ii}{i}
\newcommand{\dd}{d}
\newcommand{\vect}[1]{\boldsymbol{#1}}
\newcommand{\ie}{i.e.,}
\newcommand{\eg}{e.g.,}

\newcommand{\orcid}[1]{\href{https://orcid.org/#1}{\includegraphics[width=7pt]{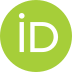}}}

\draft

\begin{document}
	
	\title{Atomic diffraction from single-photon transitions \\in gravity and Standard-Model extensions} 
	
	\author{Alexander Bott\,\orcid{0000-0002-7986-4834}}
	\email[]{alexander.bott@uni-ulm.de, alexander.bott@bott.org}
	\affiliation{Institut f{\"u}r Quantenphysik and Center for Integrated Quantum Science and Technology (IQST), Universit{\"a}t Ulm, Albert-Einstein-Allee 11, D-89081 Ulm, Germany}
	\author{Fabio Di Pumpo\,\orcid{0000-0002-6304-6183}}
	\affiliation{Institut f{\"u}r Quantenphysik and Center for Integrated Quantum Science and Technology (IQST), Universit{\"a}t Ulm, Albert-Einstein-Allee 11, D-89081 Ulm, Germany}
	\author{Enno Giese\,\orcid{0000-0002-1126-6352}}
	\affiliation{Technische Universit{\"a}t Darmstadt, Fachbereich Physik, Institut f{\"u}r Angewandte Physik, Schlossgartenstr. 7, D-64289 Darmstadt, Germany}
	\affiliation{Institut f{\"u}r Quantenoptik, Leibniz Universit{\"a}t Hannover, Welfengarten 1, D-30167 Hannover, Germany}
	
	\collaboration{This article has been published as part of the \\\emph{Large Scale Quantum Detectors Special Issue} in \href{https://doi.org/10.1116/5.0174258}{AVS Quantum Science \textbf{5}, 044402 [2023]}}
	
	\begin{abstract}
		Single-photon transitions are one of the key technologies for designing and operating very-long-baseline atom interferometers tailored for terrestrial gravitational-wave and dark-matter detection.
		Since such setups aim at the detection of relativistic and beyond-Standard-Model physics, the analysis of interferometric phases as well as of atomic diffraction must be performed to this precision and including these effects.
		In contrast, most treatments focused on idealized diffraction so far.
		Here, we study single-photon transitions, both magnetically-induced and direct ones, in gravity and Standard-Model extensions modeling dark matter as well as Einstein-equivalence-principle violations.
		We take into account relativistic effects like the coupling of internal to center-of-mass degrees of freedom, induced by the mass defect, as well as the gravitational redshift of the diffracting light pulse.
		To this end, we also include chirping of the light pulse required by terrestrial setups, as well as its associated modified momentum transfer for single-photon transitions.
	\end{abstract}

	\maketitle 
	
	\section{Introduction}
	Atomic sensors, such as clocks~\cite{Safronova2018} or atom interferometers~\cite{Bongs2019}, gain increasing attention as an alternative for detecting gravitational
	waves~\cite{Norcia2017,Tino2007,Dimopoulos2008Atomic_gravitational,Yu2011,Graham2013,Vutha2015,Kolkowitz2016,Canuel2018,Schubert2019,Badurina2020,Canuel2020,El-Neaj2020,Zhan2020,Abe2021}, possible Standard-Model violations like dark matter~\cite{Riedel2013,Derevianko2014,Arvanitaki2015,Geraci2016,Graham2016,Arvanitaki2018,Figueroa2021,Badurina2022,Du2022,Badurina2023} (DM), or violations of the Einstein equivalence principle~\cite{Damour2010,Giulini2012,Will2014,Delva2018,Asenbaum2020,Ufrecht2020,Roura2020,Lange2021,Tino2021,DiPumpo2021,DiPumpo2022,DiPumpo2023} (EEP).
	While most of these proposals rely on internal atomic transitions, \eg{} induced by (optical) single-photon transitions~\cite{Ludlow2015,Hu2017,Hu2020,Rudolph2020}, they mainly focus on the dynamics \emph{between} the times of interaction with the electromagnetic field.
	To complement these approaches, we study in this article effects of gravity, chirping, and of a basic model for DM and EEP violations on single-photon transitions, focusing on the atom's dynamics \emph{during} the interaction with electromagnetic fields.
	
	Differential measurements between two spatially separated atomic clocks or light-pulse atom interferometers, proposed for tests of gravitational waves~\cite{Dimopoulos2008Atomic_gravitational,Graham2013,Vutha2015,Kolkowitz2016} or DM~\cite{Derevianko2014,Arvanitaki2018,DiPumpo2023b,Abend2023}, probe different points in spacetime.
	While a finite propagation speed of light already contributes to the phase of a single atom interferometer, for differential setups this effects becomes even more crucial due to the large spatial separations and is in fact key to those measurement schemes.
	As a deleterious side effect, differential laser-phase noise enters the signal~\cite{Graham2013} when relying on two-photon transitions such as Raman or Bragg diffraction commonly used~\cite{Hartmann2020} for atom interferometers.
	This differential noise is suppressed using (optical) single-photon transitions, which can also drive optical atomic clocks in the Lamb-Dicke regime~\cite{Ludlow2015}.
	In addition, some atom-interferometric tests of EEP rely~\cite{Damour2010,Will2014,Delva2018,Roura2020,Lange2021,DiPumpo2022,DiPumpo2023} on such transitions.
	
	Even though some demonstrators for atom-interferometric gravitational-wave detection under construction in horizontal configurations~\cite{Canuel2018,Canuel2020} rely on Bragg diffraction~\cite{Giese2015,Hartmann2020,Dimopoulos2008Atomic_gravitational,Schubert2019}, a great number of the current vertical proposals~\cite{Badurina2020,El-Neaj2020,Abe2021} are based on single-photon transitions to avoid laser-phase noise.
	In such terrestrial setups with very long baselines~\cite{Schlippert2020,Asenbaum2020,Zhou2011,Arduini2023}, the propagation of light as well as its gravitational redshift~\cite{DiPumpo2022} have to be considered.
	Moreover, because of gravity, also chirping~\cite{Tan2017_Time} is necessary to remain resonant during the diffraction process.
	In addition, possible DM fields may further modify the transition.
	Since other long-baseline setups plan on using single-photon transitions for EEP tests~\cite{Roura2020b}, possible EEP-violating fields also have to be included for a complete description of diffraction.
	
	For the design and configuration of a differential atom-interferometric sensor, one has to decide on an atom species that has implications on atomic diffraction~\cite{Loriani20192}:
	Fermions offer weakly-allowed clock transitions, such that direct single-photon transitions without an auxiliary state are possible for fairly low laser powers.
	However, they suffer from spontaneous emissions and atom loss, resulting in an increase of shot noise and loss of coherence.
	Moreover, the increased cloud size and high expansion rate, caused by their fermionic nature, are detrimental for atom interferometers.
	Contrarily, the corresponding transitions of bosonic candidates have a clock lifetime limited by $E1-M1$ processes~\cite{Santra2004,Janson2023}, and offer the possibility to generate Bose-Einstein condensates with low expansion rate.
	However, a direct excitation of such a transition with feasible laser powers is not possible, resulting in the need for magnetically-induced transitions~\cite{Barber2006,Taichenachev2006,Hu2017,Loriani20192,Hu2020}.
	An alternative for bosons is driving the intercombination line~\cite{Rudolph2020}, although its lifetime is significantly shorter than for the clock line, limiting interrogation times and the spatial separation of atom-interferometric detectors.
	
	Consequently, this article focuses on magnetically-induced single-photon transitions.
	At the same time, by neglecting Stark shifts and by replacing the effective Rabi frequency by the corresponding actual one, our results can easily be transferred to direct single-photon transitions without magnetic fields.
	In our study, we include gravity~\cite{DiPumpo2022}, chirping~\cite{Tan2017_Time}, and a weakly-coupled, ultralight, scalar dilaton field~\cite{Damour1994,Alves2000,Damour2010,Damour2010b,Damour2012,Hees2018} as a model for both DM and EEP violations.
	We consider relativistic effects like the coupling of internal energies to the center-of-mass (c.m.) motion of the atom, induced by the mass defect~\cite{Yudin2018,Sonnleitner2018,Schwartz2019,Martinez-Lahuerta2022,Assmann2023}, which is necessary for a consistent modeling of many DM-detection schemes and EEP-test proposals.
	In Sec.~\ref{sec:MagTrans} we derive an effective two-level system for magnetically-induced single-photon transitions including all perturbations discussed above. 
	The resulting modified resonance condition, applying both to magnetically-induced and direct single-photon transitions, is discussed in Sec.~\ref{sec:ResCond}.
	We study perturbatively the time evolution during a pulse in Sec.~\ref{sec:Phaseshifts} and discuss the effects on the phase of an atom after diffraction.
	We conclude the article in Sec.~\ref{sec:concl}.
	In Appendix~\ref{sec:adel} we present technical details on adiabatic eliminations including time-dependent perturbations.
	Appendix~\ref{sec:Heisenberg} solves the time evolution in the Heisenberg picture, while Appendix~\ref{sec:phase} shows the phase resulting from the modified wave vector for non-ideal chirping.
	
	\section{Magnetically-induced single-photon transitions}
	\label{sec:MagTrans}
\begin{figure}[h]
		\centering
		\def\svgwidth{8.5cm}
		\scalebox{1}{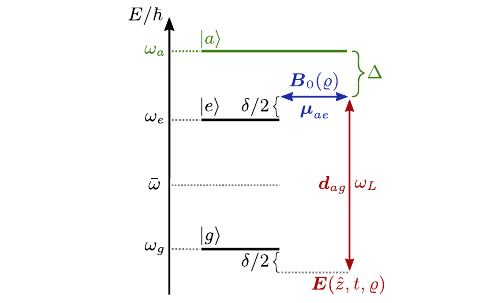}
		\caption{Term diagram of a three-level system used for magnetically-induced single-photon transitions between the ground state and the excited state.
			The system consists of ground, excited, and ancilla states $\ket{g}$, $\ket{e}$, and $\ket{a}$ at energies $\hbar \omega_{g/e/a}$, respectively.
			The mean frequency $\bar{\omega}$ of the excited and ground state has also been introduced to describe the Compton frequency of the effective two-level system.
			The ancilla state is coupled to the ground state by an electric dipole moment $\vect{d}_{ag}$ interacting with a classical electric wave $\vect{E}$.
			Its frequency $\omega_{\text{L}}$ is strongly detuned from resonance by $\Delta$.
			Furthermore, the magnetic dipole $\vect{\mu}_{ae}$ interacts with a static, real-valued magnetic field $\vect{B}_0$, coupling the ancilla state to the excite state.
			Both fields are influenced by the dilaton field $\varrho$.
			The combination of both fields drives effective transitions from the ground state to the excited state, which are detuned by $\delta$.}
		\label{fig:mSPTLevels}
	\end{figure}

\begin{figure}[h]
		\centering
		\def\svgwidth{8.5cm}
		\scalebox{1}{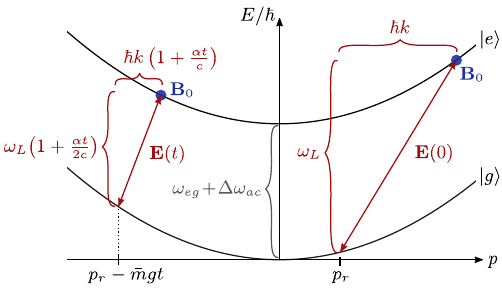}
		\caption{
			Term diagram for single-photon transitions between the effective two-level system of ground state and excited state including the kinetic contributions, where the ancilla state is not shown.
			This interaction transfers energy between the atom and the electric field $\vect{E}$, given by the field's frequency $\omega_{\text{L}}$ and momentum through its wave vector $k$. 
			The static magnetic field $\vect{B}_0$ only mediates the transition without any transfer of energy or momentum. 
			For one set of parameters, only one momentum can be resonant.
			We therefore marked the initial resonant momentum by $p_r$.
			During the interaction in gravity, the atom is accelerated to a momentum $p_r-\bar{m} g t$, which gives rise to a Doppler shift that has to be compensated to remain resonant. 
			This is done by introducing a frequency chirp $\alpha$ that modifies the resonance condition (left transition) and we find perfect compensation for $\alpha=-g$.
			However, due to the dispersion relation, the transferred momentum is modified as well in a perturbative manner.
			The energy difference between both internal states $\omega_{eg}$ is modified by a differential Stark shift $\Delta\omega_{ac}$, already included in the term diagram.
		}
		\label{fig:Chirp}
	\end{figure}

	To include physics beyond the Standard Model, we assume a classical, scalar, ultralight dilaton field $\varrho(\hat{z},t)=\varrho_{\text{DM}}(\hat{z},t)+\varrho_{\text{EP}}(\hat{z})$ which splits~\cite{DiPumpo2022} into a DM and an EEP-violating part.
	It depends in a one-dimensional model on c.m. position $\hat{z}$ and laboratory time $t$, and couples\cite{Damour1994,Damour2010} to all particles of the Standard Model including electromagnetic fields and the constituents of atoms.
	As a consequence, electric and magnetic fields as well as the mass of the atom and its internal energies depend~\cite{DiPumpo2021,DiPumpo2022} on the dilaton.
	Since we focus on magnetically-induced single-photon transitions for now, the laser does not couple the ground state $\ket{g}$ directly to the excited state $\ket{e}$ but to an ancilla state $\ket{a}$, which is strongly detuned by a frequency $\Delta$ from resonance, see Fig.~\ref{fig:mSPTLevels}.
	To transfer the population nonetheless, an additional static magnetic field, magnetically coupling the ancilla state to the excited one, is required~\cite{Barber2006,Taichenachev2006,Hu2017,Hu2020}.
	The Hamiltonian for the three-level system takes the form
	\begin{equation}
		\label{key}
		\hat{H}=\sum\limits_j\hat{H}_j\!\left(\hat{p},\hat{z},\varrho\right)\ket{j}\!\bra{j} -\hat{\!\vect{d}}\vect{E}\!\left(\hat{z},t,\varrho\right)+\hat{\!\vect{\mu}}\vect{B}_0\!\left(\varrho\right).
	\end{equation}
	It includes the electric dipole operator $\hat{\!\vect{d}} = \vect{d}_{ag}\ket{a}\!\bra{g}+\text{h.c.}$ interacting with a classical electric wave $\vect{E}$, as well as the magnetic dipole operator $\hat{\!\vect{\mu}} = \vect{\mu}_{ae}\ket{a}\!\bra{e}+\text{h.c.}$ interacting with a static, real-valued magnetic field $\vect{B}_0$.
	Here, the dipole transition element $\vect{d}_{ag}$ is coupling the ground state to the ancilla state, while the magnetic dipole transition element $\vect{\mu}_{ae}$ couples the excited state to the ancilla state.
	The c.m. momentum $\hat{p}$ and position $\hat{z}$ operators point in the $z$-direction of the laboratory frame, with $[\hat{z}, \hat{p}]= \ii \hbar$.
	The time- and position-dependent dilaton field $\varrho(\hat{z},t)$ is specified later.
	Since it couples to the mass-energy of the atom~\cite{Yudin2018,Sonnleitner2018,Schwartz2019,Martinez-Lahuerta2022,Assmann2023}, it also influences the motion of the atom in each internal state $\ket{j}$ through
	\begin{equation}
		\label{Hj}
		\hat{H}_j\!\left(\hat{p},\hat{z},\varrho\right) = m_j(\varrho)c^2+\frac{\hat{p}^2}{2m_j(\varrho)}+m_j(\varrho)g\hat{z},
	\end{equation}
	with $c$ being the speed of light and the gravitational acceleration $g$ aligned with the $z$-direction.
	We emphasize that the state-dependent mass $m_j(\varrho)$ depends not only on the dilaton but also on the atomic state $\ket{j}$, \ie{} it includes the mass defect~\cite{Yudin2018,Sonnleitner2018,Schwartz2019,Martinez-Lahuerta2022,Assmann2023}, and introduces a coupling of internal states to c.m. operators.
	Further, we assume
	\begin{equation}
		\label{key}
		\vect{E}\!\left(\hat{z},t,\varrho\right) = \vect{E}_0(\varrho)\,\ee^{\ii\varphi_\text{L}(\hat{z},t)}+\text{ h.\,c. }
	\end{equation}
	for the electric wave propagating against gravity with dilaton-dependent amplitude $\vect{E}_0(\varrho)$ and phase
	\begin{equation}
		\label{eq:PhaseLightField}
		\varphi_\text{L}(\hat{z},t) = k \hat{z} \left( 1 + \frac{\alpha t}{c} - \frac{(g+\alpha) \hat{z}}{2c^2}\right)- \varphi_0 - \omega_\text{L}t \left( 1 + \frac{\alpha t}{2c} \right)
	\end{equation}
	Here, $\varphi_0$ is the phase offset of the laser, while $\omega_\text{L}$ is the frequency of the electric field and $\alpha$ is its chirp rate in units of an acceleration. 
	Chirping is necessary to compensate for Doppler shifts and induce resonant transitions in gravity, as shown in Fig.~\ref{fig:Chirp}, and therefore of particular relevance for vertical setups.
	The first term in Eq.~\eqref{eq:PhaseLightField} is the spatial mode function and gives rise to the momentum imparted on the atom via a displacement operator.
	Without taking into account any perturbations, it transfers a momentum $\hbar k$ to the atom, where the dispersion relation $kc=\omega_\text{L}$ holds, in contrast to two-photon processes where the effective wave vector can be tuned independently from the transferred energy.
	This first term also includes a time-dependent modification from chirping, usually not present for two-photon transitions, as well as a gravitational redshift factor $g\hat{z}^2/(2c^{2})$ from the gravitational modification of the wave vector.
	Since chirping acts as an additional acceleration, it also appears as an accelerational redshift factor $\alpha\hat{z}^2/(2c^{2})$.
	Hence, the momentum transfer does not include a redshift modification proportional to $\hat{z}^2$ for perfect chirping $\alpha=-g$. 
	These perturbations can be derived from a gravitationally modified eikonal equation~\cite{Bartelmann2019}. 
	
	\subsection{Rotating-wave approximation}
	To remove the time dependence from the phase of the light field, we change into a co-rotating picture by applying the unitary transformation
	\begin{align}
		\label{key}
		\begin{split}
			\hat{U}_{\text{rot}} =& \left(\ket{a}\!\bra{a}+\ket{e}\!\bra{e}\right)\ee^{\ii[\varphi_\text{L}(\hat{z},t)/2-\bar{\omega}t]}\\&+\ket{g}\!\bra{g}\ee^{-\ii[\varphi_\text{L}(\hat{z},t)/2+\bar{\omega}t]}
		\end{split}
	\end{align}
	with mean frequency $\bar \omega = (\omega_e+\omega_g)/2$ between ground and excited state, see Fig.~\ref{fig:mSPTLevels}, where $\omega_j = m_j(0)c^2/\hbar$ is connected to the mass $m_j(0)$ in state $\ket{j}$ evaluated at the Standard-Model value for $\varrho=0$.
	The transformation removes the laser phase and includes a momentum displacement, since because of momentum conservation atoms that have absorbed light from the electric field experience a momentum kick.
	For complex dipole moments $\vect{d}_{ag}$ and $\vect{\mu}_{ae}$ the transformation can be modified to remove respective phases.
	Since we consider a static magnetic field, neither energy nor momentum are transferred by this field, as apparent from Fig.~\ref{fig:Chirp}, and no oscillating phase factors appear.
	Moreover, the product $\hat{\!\vect{\mu}}\vect{B}_0\!(\varrho)$ remains unchanged by the transformation such that no corresponding transformation for the magnetic field is necessary.
	In contrast, we find four contributions from $-\hat{\!\vect{d}}\vect{E}\!\left(\hat{z},t,\varrho\right)$.
	After the transformation described above, two counterrotating terms arise that oscillate with twice the (optical) laser frequency $\omega_\text{L}$.
	Hence, we perform the rotating-wave approximation~\cite{Sanz2016} by neglecting these quickly oscillating terms.
	The validity of this approximation is independent of the internal energies, instead a large temporal derivative of the laser phase $\left|\left|\hbar\dot  \varphi_\text{L}/ [\vect{d}_{ag}\vect{E}_0(\varrho)]\right|\right|\gg 1$ is required.
	
	After the rotating-wave approximation, we find the Hamiltonian in the rotating picture and cast it into a matrix form.
	To this end, we choose the basis $\left(\ket{\psi_a},\ket{\psi_e},\ket{\psi_g}\right)^\text{T}$, where $\ket{\psi_j}$ describes the c.m. motion in state $\ket{j}$.
	We arrive at the Hamiltonian
	\begin{equation}
		\label{3levelHam}
		\hat{H}_{\text{rot}}
		= \frac{\hbar}{2} \begingroup
		\begin{pmatrix} 2[\hat{\nu}_{a}+\Delta] & \Omega_B(\varrho) & \Omega_E(\varrho) \\
			\Omega_B(\varrho) &2 \hat{\nu}_{e}+\delta & 0 \\ 
			\Omega_E(\varrho) & 0 &2\hat{\nu}_{g} - \delta 
		\end{pmatrix}\endgroup = \hbar \begingroup
		\begin{pmatrix} \hat{\nu}_{a}+\Delta & \hat{T}^\dagger \\
			\hat{T} & \hat{\nu}_{eg} 
		\end{pmatrix}\endgroup
	\end{equation}
	with auxiliary detuning $\Delta = \omega_a-\bar\omega-\omega_\text{L}/2$, two-level detuning $\delta = \omega_e-\omega_g - \omega_\text{L}$, both depicted in Fig.~\ref{fig:mSPTLevels}, as well as the Rabi frequencies $\Omega_E(\varrho) = -2\vect{d}_{ag}\vect{E}_0(\varrho)/\hbar$ and $\Omega_B(\varrho) = 2\vect{\mu}_{ae}\vect{B}_0(\varrho)/\hbar$ for the electric and magnetic transition, respectively.
	Here, the c.m. dynamics associated with different internal states via the mass defect is given by
	\begin{subequations}
		\begin{align}
			\hat{\nu}_{a/e} &= \hat{H}_{a/e}\!\left(\hat{p}+\hbar \hat{\kappa},\hat{z},\varrho\right)/\hbar-\omega_{a/e}+\frac{k \alpha (\hat{z}-ct)}{2c}\\
			\hat{\nu}_{g} &= \hat{H}_{g}\!\left(\hat{p}-\hbar \hat{\kappa},\hat{z},\varrho\right)/\hbar-\omega_{g}-\frac{k \alpha (\hat{z}-ct)}{2c}
		\end{align}
	\end{subequations}
	with the momentum displacement $\hbar \hat{\kappa} =\hbar k\left[1 +\alpha t/c-(g+\alpha)\hat{z}/c^2\right]/2$.
	Moreover, we defined in a second step in Eq.~\eqref{3levelHam} the abbreviations
	\begin{align}
		&\hat{T} = \frac{1}{2}\renewcommand*{\arraystretch}{1.5}\begin{pmatrix} \Omega_B(\varrho) \\ \Omega_E(\varrho) \end{pmatrix} \text{ and }
		\hat{\nu}_{eg} =  \renewcommand*{\arraystretch}{1.5}\begin{pmatrix} \hat{\nu}_e+\delta/2 & 0 \\ 0  & \hat{\nu}_g-\delta/2 \end{pmatrix}
	\end{align}
	for a more compact notation.
	For strongly detuned ancilla states where $\Delta$ is much larger than all other frequency scales, one expects that this state is only virtually populated for short times.
	Similar to two-photon transitions, we can adiabatically eliminate this state and describe the effective dynamics only between ground and excited state induced by single-photon transitions.
	
	\subsection{Adiabatic elimination}
	To reduce the Hilbert space from three to two states and to obtain an effective two-level system, we define~\cite{Sanz2016} a projector $\hat{P}$ with $\ket{\psi_a} = \hat{P} \bigl(\ket{\psi_e},\ket{\psi_g}\bigr)^T$.
	Since $\hat{H}_{\text{rot}}$ depends on time, $\hat{P}$ may also be time dependent, in contrast to the conventional treatment~\cite{Sanz2016,Paulisch2014,Brion2007}.
	In Appendix~\ref{sec:adel}, we derive the Schr\"odinger equations for $\ket{\psi_a}$ and $\bigl(\ket{\psi_e},\ket{\psi_g}\bigr)$, and find the Bloch equation
	\begin{equation}\label{Bloch}
		\Delta \hat{P}+\hat{T}^\dagger = -\hat{\nu}_a\hat{P}+\hat{P}\hat{\nu}_{eg}+\ii\frac{\partial}{\partial t}\hat{P}+\hat{P}\hat{T}\hat{P},
	\end{equation}
	taking into account a possible time dependence of the projector.
	Afterwards, we make a series ansatz $\hat{P} = \sum_{n=0}^{\infty}\hat{P}_n$ for the projector, assuming a perturbative expansion obeying $||\hat{P}_{n+1}||\ll||\hat{P}_n||$.
	We also require that $\Delta$ is much larger than all other frequency scales of the Hamiltonian, leading to expansion parameters $||\hat{\nu}_{a}/\Delta||$, $||\hat{\nu}_{eg}/\Delta||$, and $||\hat{T}/\Delta||$.
	Furthermore, we assume that the frequency scales introduced by chirping and the dilaton are also small compared to $\Delta$, \ie{} $||\dot{\hat{\nu}}_{a}/\Delta^2||$, $||\dot{\hat{\nu}}_{eg}/\Delta^2||$, and $|| \dot{\hat{T}}/\Delta^2||$ are of even higher order.
	These relations imply that $||\dot{\hat{P}}_{n}/\Delta||$ is also of higher order.
	
	Next, we search for a relation $||\hat{P}_{n+1}||\sim||\hat{P}_n/\Delta||$ which allows us to truncate the series.
	If $||\hat{P}_0||\sim | \Delta^{-1}|$ holds, the right-hand side of the Bloch equation is smaller than the left-hand side by at least one order in the expansion parameters.
	As a consequence, we neglect terms on the right, resulting in the lowest order $\hat{P}_0=-\hat{T}^\dagger/\Delta$, fulfilling the condition $||\hat{P}_0 ||\sim |\Delta^{-1}|$.
	Higher-order terms and a more detailed derivation of the Bloch equation can be found in Appendix~\ref{sec:adel}.
	Here, we use only the lowest-order contributions to determine the Hamiltonian acting on the reduced space containing only excited and ground state, \ie{} $\hat{\mathcal{H}}_{\text{rot}} /\hbar \cong-\hat{T}\hat{T}^\dagger/\Delta +\hat{\nu}_{eg}$.
	By assuming a weakly-coupling, ultralight dilaton field with $||\varrho(\hat{z},t)||\ll 1$, such that $\partial \Omega_{E/B} /\partial \varrho |_{\varrho=0} \leq \Omega_{E/B}(0)$, we can neglect terms scaling with $\varrho(\hat{z},t)/\Delta$ in the Hamiltonian.
	This way, we obtain an effective two-level Hamiltonian in matrix form
	\begin{equation}
		\label{keyHam}
		\hat{\mathcal{H}}_{\text{rot}} = \frac{\hbar}{2}\renewcommand*{\arraystretch}{2}
		\begin{pmatrix}
			2\hat{\bar{\nu}}+\hat{\nu} & \Omega\\
			\Omega & 2\hat{\bar{\nu}}-\hat{\nu}
		\end{pmatrix}
	\end{equation}
	with effective Rabi frequency $\Omega = -\Omega_{B}(0)\Omega_{E}(0)/(2\Delta)$, where all dilaton perturbations of the electromagnetic fields can be neglected.
	The remaining influence of the dilaton field is restricted to the diagonal elements of the Hamiltonian, which includes the mean energy $\hat{\bar{\nu}}=(\hat{\nu}_e+\hat{\nu}_g)/2+\bar{\omega}_{ac}$ and effective detuning $\hat{\nu}=\hat{\nu}_e-\hat{\nu}_g+\delta+\Delta\omega_{ac}$ of the effective two-level system.
	This effective two-level description introduces mean and differential Stark shifts, given by $\bar{\omega}_{ac} = -\left[\Omega_E^2(0)+\Omega_B^2(0)\right]/(8\Delta)$ and $\Delta\omega_{ac} = \left[\Omega_E^2(0)-\Omega_B^2(0)\right]/(4\Delta)$, respectively.
	
	Note that for direct single-photon transitions without a magnetic field, one can directly use a two-level Hamiltonian in the form of Eq.~\eqref{keyHam} by assuming vanishing Stark shifts and replacing the effective Rabi frequency by the fundamental one.
	All other features, in particular the resonance condition discussed in Sec.~\ref{sec:ResCond}, remain of the same form.
	
	\section{Resonance condition for single-photon transitions}
	\label{sec:ResCond}
	For discussing single-photon transitions, both magnetically-induced and direct ones, one needs to analyze the resonance condition based on the (effective) Hamiltonian $\hat{\mathcal{H}}_{\text{rot}}$, including chirping and perturbations.
	To this end, we split the dilaton field $\varrho(\hat{z},t)=\varrho_{\text{DM}}(\hat{z},t)+\varrho_{\text{EP}}(\hat{z})$ into DM and EEP-violating contributions.
	We consider an ultralight dilaton field expanded locally in the laboratory frame.
	Cosmological and galactic contributions to this field act as background resulting in DM, and local gravitational contributions sourced by Earth act as EEP violations~\cite{DiPumpo2022}.
	The DM part $\varrho_{\text{DM}}(\hat{z},t)=\bar{\varrho}_0\cos(\omega_{\varrho}t-k_{\varrho}\hat{z}+\phi_{\varrho})$ includes the perturbative amplitude $\bar{\varrho}_0$, the frequency $\omega_{\varrho}$, the wave vector $k_{\varrho}$, and the phase $\phi_{\varrho}$ of the DM field.
	The EEP-violating part $\varrho_{\text{EP}}(\hat{z}) = \beta_S g\hat{z}c^{-2}$ contains the coefficient $\beta_S$ that arises from the expansion of a source mass, \eg{} Earth, in orders of the dilaton field.
	Similarly, we expand the internal-state-dependent mass of the atom $m_j(\varrho) = m_j(0)[1+\beta_j\varrho(\hat{z},t)]$ around its Standard-Model value $ m_j(0)$, with linear expansion coefficient $\beta_j$.
	We define the mass defect~\cite{Assmann2023} as $m_{e/g}(0) = \bar{m}\pm\hbar\omega_{eg}/(2c^2)$, with mean mass $\bar{m}$ and energy difference between the internal states $\omega_{eg}=\omega_e-\omega_g$.
	Then, we insert the dilaton field and the mass defect into Eq.~\eqref{Hj} and approximate all terms to first order in $c^{-2}$, $\bar{\varrho}_0$, $\beta_{e/g}$, and $\omega_{eg}/\bar{\omega}$.
	Consequently, DM introduces time-dependent oscillations of the internal energies~\cite{Derr2023}, while the transferred momentum is modified according to Eq.~\eqref{eq:PhaseLightField} by chirping~\cite{Tan2017_Time} and by the gravitational redshift~\cite{DiPumpo2022} of the wave vector $k$.
	The kinetic and potential energies include state-dependent modifications due to the mass defect and possible EEP violations.
	
	Then, we remove the unperturbed free-fall part, the recoil frequency $\omega_k = \hbar k^2/(2\bar{m})$, and other constant energy shifts by using the transformation into a mean Heisenberg picture 
	\begin{equation}
		\label{key}
		\hat{U}^{\text{H}}=\exp\left[-\ii\left(\frac{\hat{p}^2}{2\bar{m} \hbar }+\frac{\bar{m}g\hat{z}}{\hbar}+\frac{\omega_k  }{4}+\bar{\omega}_{ac}\right)t\right].
	\end{equation}
	For the Heisenberg trajectories, we find the replacements $\hat{p}\rightarrow\hat{p}_{\text{H}} = \hat{p}-\bar{m}gt$ and $\hat{z}\rightarrow\hat{z}_{\text{H}} = \hat{z}+\hat{p}t/\bar{m}-gt^2/2$, which correspond to classical trajectories in an unperturbed gravitational potential.
	The transformation leads to the (effective) detuning in the Heisenberg picture
	\begin{align}
		\label{eq:DetInHeis}
		\begin{split}
			\hat{\nu}_{\text{H}} =& (\omega_{eg}+\hat{\nu}_k+\Delta\omega_{ac})-\omega_\text{L}-k(\alpha+g)t+\frac{k\alpha\hat{z}_{\text{H}}}{c}\\&+\frac{\omega_{eg}}{\bar{\omega}}\left(\frac{\bar{m}g\hat{z}_{\text{H}}}{\hbar}-\frac{\hat{p}_{\text{H}}^2}{2\bar{m}\hbar}-\frac{\omega_k}{4}\right)+\bar{\omega}\Delta\beta\varrho(\hat{z}_{\text{H}},t)\\
			&+\left\{\frac{k\hat{p}_{\text{H}}}{2\bar{m}},\frac{\alpha t}{c}-\frac{(g+\alpha)\hat{z}_{\text{H}}}{c^2}\right\}
		\end{split}
	\end{align}
	with $\Delta \beta = \beta_e-\beta_g$ and where we introduced the anti-commutator $\{\hat{A}, \hat{B}\}=\hat{A}\hat{B}+\hat{B}\hat{A}$ for two operators $\hat{A}$ and $\hat{B}$.
	The first parenthesis in Eq.~\eqref{eq:DetInHeis} describes the energy difference between both internal states, including the Doppler frequency $\hat{\nu}_k = k\hat{p}/\bar{m}$ associated with the recoil during absorption, as well as the energy shift $\Delta \omega_{ac}$ caused by Stark shifts.
	This energy has to be transferred by the light field, \ie{} by $\omega_\text{L}$.
	The c.m. motion of the atom during the interaction introduces a Doppler shift $-kgt$ that can be compensated by the term $-k\alpha t$ if the laser frequency is appropriately chirped.
	
\begin{figure}[h]
		\centering
		\def\svgwidth{8.5cm}
		\scalebox{1}{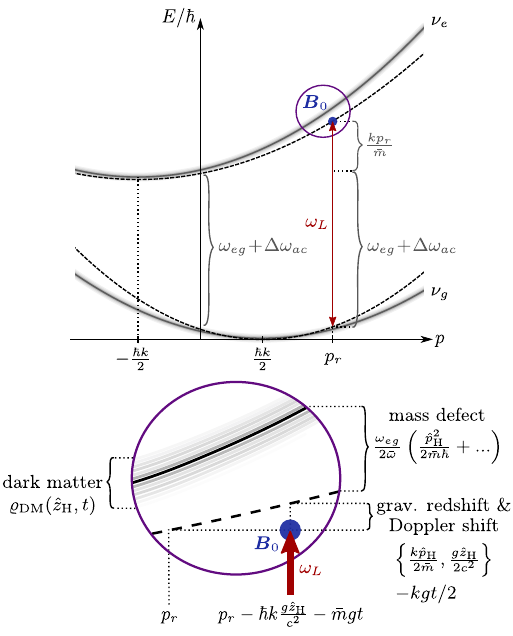}
		\caption{
			Perturbations to the resonant transition in an energy-momentum diagram shifted in the co-rotating frame.
			The momentum of the excited state is displaced by $-\hbar k/2$ in the co-rotating frame, the one of the ground state by $+\hbar k/2$.
			The unperturbed energies are given by the dashed parabolas, but are modified to $\hbar \nu_{e/g}$ (solid lines) including all perturbations.
			Without any perturbations, the momentum $p_r$ is resonant and the resonance condition includes the initial Doppler detuning $kp_r/\bar m$ as well as the energy difference $\omega_{eg}$ and differential Stark shift $\Delta \omega_{ac}$. 
			This resonance is modified by perturbations, as shown in the magnified part below the figure.
			Gravity causes an additional Doppler shift as shown in Fig.~\ref{fig:Chirp}, as well as a gravitational redshift, which shifts the momentum transferred by the electric field.
			Both effects can be compensated by chirping the frequency of the electric field with chirp rate $\alpha=-g$.
			The mass defect shifts the internal energies depending on the atom's momentum, modifying the dispersion relation.
			Finally, dark matter causes the energies of the internal states to oscillate in time, as highlighted by the blurred line of the internal energies.
		}
		\label{fig:modResCon}
	\end{figure}

	However, the momentum transfer is modified by chirping so that the current c.m. position of the atom enters.
	Additional position- and time-dependent modifications of internal, potential, and kinetic energies, as well as the recoil frequency caused by the mass defect, the coupling to DM, and EEP violations appear. 
	Further modifications to the Doppler frequency arise due to the modification of the momentum transfer, as can be seen from the anti-commutator in the last line of Eq.~\eqref{eq:DetInHeis}. 
	All of these modifications cannot be included into the static frequency $\omega_\text{L}$ in actual experimental setups.
	Hence, we choose the laser frequency $\omega_\text{L} = \omega_{eg}+k p_\text{r}/\bar{m}+\Delta\omega_{ac}$ in the Heisenberg picture, representing the unperturbed resonance condition but not obeying the modified resonance condition due to perturbations.
	The second contribution in the unperturbed resonance condition describes the Doppler shift at the beginning of the interaction and defines the resonant momentum $p_\text{r}$.
	In this case we obtain
	\begin{align}
		\begin{split}
			\hat{\nu}_{\text{H}} =& \frac{k\left(\hat{p}-p_\text{r}\right)}{\bar{m}}-k(\alpha+g)t+\frac{k\alpha\hat{z}_{\text{H}}}{c}\\
			&+\frac{\omega_{eg}}{\bar{\omega}}\left(-\frac{\hat{p}_{\text{H}}^2}{2\bar{m}\hbar}+\frac{\bar{m}g\hat{z}_{\text{H}}}{\hbar}-\frac{\omega_k}{4}\right)+\bar{\omega}\Delta\beta\varrho(\hat{z}_{\text{H}},t)\\
			&+\left\{\frac{k\hat{p}_{\text{H}}}{2\bar{m}},\frac{\alpha t}{c}-\frac{(g+\alpha)\hat{z}_{\text{H}}}{c^2}\right\},
		\end{split}
	\end{align}
	which depends only on chirping, gravity, and velocity selectivity $(\hat{p}-p_\text{r})/\bar m$, \ie{} a deviation of the momentum distribution from the resonant momentum, as well as perturbations induced by the dilaton field, the mass defect and the modified momentum transfer.
	We illustrate this modified resonance condition given by $\hat{\nu}_{\text{H}}$ in Fig.~\ref{fig:modResCon}.
	For the reduced mean energy in the Heisenberg picture we find 
	\begin{equation}\label{nuHbar}
		\hat{\bar{\nu}}_{\text{H}} = \bar{\omega}\bar{\beta}\varrho(\hat{z}_{\text{H}},t)-\frac{k\hat{p}_{\text{H}}}{4\bar{m}}\frac{\omega_{eg}}{\bar{\omega}}+\frac{\omega_k}{2}\left(\frac{\alpha t}{c}+\frac{\alpha^2 t^2/2-(g+\alpha)\hat{z}_{\text{H}}}{c^2}\right)
	\end{equation}
	with $\bar{\beta} = (\beta_e+\beta_g)/2$.
	
	We observe that modifications to Doppler shifts arising from the mass defect and the modified momentum transfer cannot be fully compensated.
	In summary we find from the effective resonance condition that even in the case $\alpha=-g$ and neglecting velocity selectivity, perturbative effects persist.
	As such, it causes a different time evolution during the pulse compared to the unperturbed case, and, thus, leads to phase shifts on diffracted atomic wave packets that scale with the pulse duration.
	We stress that these results are valid for both magnetically-induced single-photon transitions and, with the replacements discussed above, also for direct ones.
	
	\section{Phase shifts between diffracted wave packets}
	\label{sec:Phaseshifts}
	Since the modified resonance condition affects the time evolution during the pulse, perturbations lead to additional phase shifts between diffracted and undiffracted atoms.
	In our study, we included as perturbations possible EEP violations, DM, chirping, the redshifted momentum transfer, and the mass defect.
	In the following, we analyze effects of these perturbations on the phase imprinted on atoms by the pulses.
	
	We assume that the unperturbed resonance condition is sufficiently fulfilled, so that the residual detuning $\hat{\nu}_\text{H}$ and mean energy $\hat{\bar \nu}_\text{H}$ can be used as perturbative quantities, leading to a Rabi frequency that is dominant compared to these frequency scales, \ie{} $\Omega\gg ||\hat{\nu}_\text{H}||,||\hat{\bar \nu}_\text{H}||$.
	Similar to previous works~\cite{Stoner2011,Bertoldi2019}, we change into an interaction picture with respect to the unperturbed Rabi oscillation via the transformation $\hat{U}_{\varphi}$, and then solve the time evolution of the remaining perturbations with a Dyson-series approach, see Appendix~\ref{sec:Heisenberg} for details.
	Since we consider weakly-coupled, ultralight DM, we expect it to have a small Compton frequency and associated  wave vector~\cite{Arvanitaki2015}.
	Therefore, we neglect its time dependence over the duration of a pulse and evaluate the field at the initial time of the pulse.
	This assumption results in the relation $\varrho_{\text{DM}}(\hat{z}_{\text{H}},t)\cong\varrho_{\text{DM}}(\hat{z},0)$, with perturbative parameters $\omega_{\varrho}/\Omega$, $k_{\varrho}\hat{p}/(\bar{m}\Omega)$, and $k_{\varrho}g/\Omega^2\ll1$.
	With this procedure, we arrive at the solution $\hat{U}$ for the time evolution in the Heisenberg picture, see Appendix~\ref{sec:Heisenberg}.
	Transformed back into the Schr\"odinger picture, we find the time evolution
	\begin{align}\label{US}
		\begin{split}
			\hat{U}^{\text{S}}(t) =& \hat{U}_{\text{rot}}(t)\hat{U}^{\text{H}}(t)\hat{U}(t)\hat{U}_{\text{rot}}^{\dagger}(0)=\sum\limits_{j,n=e,g}{\hat{U}^{\text{S}}_{n,j}\ket{n}\!\bra{j}}.
		\end{split}
	\end{align}
	All time-dependent operators $\hat{U}^{\text{S}}_{n,j}$, acting on the c.m. motion and describing the transition between internal states, are listed in Appendix~\ref{sec:Heisenberg}.
	For simplicity and clarity, we discuss effects of mirror pulses in the following, being specified by $\Omega t = \pi$.
	We consider a diffracted atomic wave function in momentum representation
	\begin{equation}
		\psi_{n,j}(p,t) = \bra{p}\hat{U}^{\text{S}}_{n,j}\ket{\psi_{0,j}},
	\end{equation}
	which depends on the initial internal state $\ket{j}$ and the final state $\ket{n}$.
	The momentum of atoms initially in the ground state is increased by $\hbar k$ after diffraction.
	Hence, we find the phase difference
	\begin{align}
		\begin{split}
			\phi_{\pi} =& \arg\Bigl[\psi_{e,g }(p+\hbar k,t)\Bigr]-\arg\Bigl[\psi_{g,e}(p,t)\Bigr]\\
			=& \phi_0+\phi_{\text{DM}}+\phi_{\text{EP}}+\phi_{\text{MD}}+\phi_{\text{WV}}
		\end{split}
	\end{align}
	between phases of atoms after their transition from ground to excited state and the vice versa case, which is the relevant contribution for a mirror pulse in an atom interferometer.
	
	In the following, we list phase contributions to $\phi_{\pi}$ of the unperturbed part, DM, EEP violations, the mass defect, and the wave vector modification caused by gravity and chirping.
	To this end, we assume a Gaussian initial state for the c.m. wave packet before the pulse in momentum representation 
	\begin{equation}
		\braket{p |\psi_{0,e/g}} \propto \exp\left( -\frac{(p-p_{e/g})^2}{2 \hbar^2\sigma_{e/g}^2}+\ii \frac{p z_{e/g}}{\hbar}\right),
	\end{equation}
	with internal-state-dependent widths $\hbar \sigma_{e/g}$, momentum expectation values $p_{e/g}$, and mean positions $z_{e/g}$.
	\begin{subequations}
		\label{eq.phase_contributions}
		
		This wave packet results to leading order in the unperturbed phase
		\begin{align}
			\begin{split}
				\phi_0 =& - 2 \varphi_0 - k(\bar{z}+ \Delta z /2) -\Delta z\frac{p+\bar{m}gt}{\hbar} \\
				& + 2k\frac{g+\alpha}{\Omega^2}- \omega_\text{L} \left( t + \frac{\alpha t^2}{2c} \right) 
			\end{split}
		\end{align}
		which contains the difference of initial position $\Delta z = z_e-z_g$ and the mean initial position $\bar{z}=(z_e+z_g)/2$ .
		This unperturbed part includes twice the laser phase offset $\varphi_0$ and a contribution $k \bar z $ that arises from the unperturbed momentum transfer.
		Moreover, separation terms that depend on $\Delta z$ and the momentum appear.
		Another term results from a chirping mismatch $g+\alpha$, but is suppressed by the square of the Rabi frequency.
		The unperturbed part also includes the laser frequency during the interaction because of the internal transitions, similar to atomic clocks when including chirping.
		Besides, for $\alpha=-g$ we would obtain a $kgt^2/2$ term from the laser frequency.
		
		The phase difference induced by DM during the pulse
		\begin{align}
			\label{eq.phi_dm}
			\begin{split}
				\phi_{\text{DM}}\cong \bar{\omega}t\bar{\beta}\left.\frac{\partial}{\partial z}\varrho_{\text{DM}}(z,0)\right\vert_{z=\bar{z}}k_{\varrho}\Delta z
			\end{split}
		\end{align}
		results from assuming $ k_{\varrho}\ll\hbar \sigma_{e/g}$, \ie{} taking into account only the mean coupling of DM to the rest mass of the atom, depending on the residual phase of the DM wave $\phi_{\varrho}-k_{\varrho}\bar z$.
		
		Possible EEP violations lead to the phase difference
		\begin{equation}
			\label{eq.phi_ep}
			\phi_{\text{EP}}=-\bar{\beta}\beta_S\frac{\bar{m}g\Delta z}{\hbar}t-2\Delta\beta\beta_S\frac{g / v_\text{r}}{\Omega^2}\left(\frac{kp}{\bar{m}}+\omega_k+\frac{kgt}{2}\right),
		\end{equation}
		with recoil velocity $v_\text{r}=\hbar k / \bar m$ , Doppler frequency  $k p/\bar{m}$, and recoil frequency $\omega_k = \hbar k^2/(2 \bar m )$.
		The first contribution stems from different initial positions in the mean gravitational potential, while the second contribution originates from different gravitational accelerations for each internal state, resulting in different Doppler shifts.
		
		The mass defect causes a phase difference between excited and ground state
		\begin{equation}\label{key}
			\phi_{\text{MD}}=-4\frac{\omega_{eg}}{\bar{\omega}}\frac{g / v_\text{r}}{\Omega^2}\left(\frac{kp}{\bar{m}}+\omega_k+\frac{kgt}{2}\right)
		\end{equation}
		resulting from asymmetric diffraction between those two states under gravity.
		
		Finally, the wave-vector modification leads for perfect chirping, \ie{} $\alpha=-g$, to the phase difference
		\begin{equation}
			\label{eq:PhaseWaveVEc}
			\phi_{\text{WV}}=\frac{kg t}{\Omega} \left[\frac{\Omega}{c}
			\left(\bar{z}-\frac{p t}{\bar m}-\frac{v_\text{r} t}{2}-\frac{gt^2}{2}\right)+\frac{2p}{\pi \bar{m}c}+\frac{v_\text{r}}{\pi c}+\frac{gt}{\pi c}\right]
		\end{equation}
	\end{subequations}
	where chirping has modified the wave vector and, by that, the momentum transfer during the pulse.
	Additional contributions for imperfect chirping $\alpha \neq -g $ are listed in Appendix~\ref{sec:phase}, and only in this generalized case the finite widths $\hbar \sigma_{e/g}$ appear in the phase difference.
	In this case, additional contributions due to the gravitational modification of the wave vector of light arise.
	Note that in counterpropagating two-photon transitions, such effects cannot be compensated by chirping, in contrast to single-photon transitions.
	
	Among the phases listed above, the dominant phase contribution stems from the unperturbed single-photon phase $\phi_0$, even though some of its terms vanish for perfect chirping.
	Since we observe
	\begin{equation}
		\phi_\text{WV}= \phi_\text{MD} \frac{k c}{2 \omega_{eg}} \left[1+\frac{\pi^2}{2} -  \frac{\bar z \pi^2}{ 2 pt/\bar m +v_\text{r} t + g t^2 }\right],
	\end{equation}
	we conclude that the relativistic contributions induced by a wave-vector modification and by the mass defect are of similar order, but their relative size depends on the duration of the pulse.
	In fact, for short pulses one expects  $\phi_{\text{WV}}$ to dominate.
	
	In an atom interferometer that tests for dark matter or EEP violations, one seeks to bound the coupling parameters $\bar{\beta}$, $\Delta \beta$, or $\beta_S$.
	While a measurement of the phases $\phi_{\text{EP}}$ and $\phi_{\text{DM}}$ may serve such a purpose, stricter bounds are given by the interferometer signal induced during propagation in between the diffracting pulses. 
	Since the phases presented above are on the timescale of the pulse duration $t\sim 1/\Omega$ that is much smaller than the interrogation time of the interferometer, we refrain from discussing the bounds implied by Eqs.~\eqref{eq.phi_dm} and~\eqref{eq.phi_ep} or their relative size.
	
	Similar results for beam splitters with $\Omega t = \pi/2$ can be obtained from the time evolution in Appendix~\ref{sec:Heisenberg}.
	Since the effects of the perturbations presented here take place \emph{during} the pulse, they will be suppressed for short durations~\cite{Bertoldi2019}.
	Such short pulse durations are limited by the effective Rabi frequency, but in contrast to Bragg diffraction~\cite{Hartmann2020} no higher diffraction orders arise.
	Hence, in principle a short pulse duration and a large effective Rabi frequency are possible, but they are limited by the coupling strength and the available laser power and magnetic field intensity.
	
	For direct single-photon transitions without magnetic field~\cite{Rudolph2020}, the Rabi frequency would be time- and position-dependent due to dilaton modifications, which are suppressed by $\varrho/\Delta$ in the magnetically induced case.
	However, to first order in $\bar{\varrho}_0$ and $c^{-2}$, the Rabi frequency commutes with $\hat{\nu}_\text{H}$ and $\hat{\bar{\nu}}_\text{H}$ even without magnetic field, so that our results immediately transfer to direct single-photon transitions.
	
	\section{Conclusions}
	\label{sec:concl}
	In this article, we described the light-matter interaction for single-photon transitions, including gravity and perturbations like DM, EEP violations, as well as a modified wave vector of light caused by gravity and chirping.
	As such, we derived an effective two-level system for single-photon transitions with a magnetic field.
	Considering the dynamics during the pulse shows that the resonance condition contains modifications from these perturbations, causing imperfect diffraction of atoms.
	Moreover, we showed that these perturbations result in additional phase contributions for atoms after diffraction.
	For example, in contrast to two-photon diffraction, chirping leads to a modified momentum transfer due to a fixed dispersion relation, which would even be present in the instantaneous-pulse approximation.
	In contrast, the gravitational redshift of light implies a modified momentum transfer for two-photon transitions~\cite{DiPumpo2022}.
	As we demonstrated, however, this effect is suppressed in single-photon transitions for perfect chirping.
	
	We emphasize that our results can be immediately transferred to direct single-photon transitions without a magnetic field, starting directly from a two-level system and not an effective one.
	Then, effective quantities like the Rabi frequency are replaced by their counterparts, while no adiabatic elimination is necessary.
	
	Vertical terrestrial proposals~\cite{Badurina2020,El-Neaj2020,Zhan2020,Abe2021,Arduini2023,DiPumpo2023b,Abend2023} take place in gravity.
	Hence, chirping is crucial to stay on resonance, which becomes particularly relevant for very-long-baseline setups~\cite{Schlippert2020,Asenbaum2020,Zhou2011}.
	As discussed here, chirping leads to a modified momentum transfer, which has not been considered in earlier works so far.
	Hence, by including all relevant perturbations consistently, we have performed the first step by focusing on the dynamics during the pulse.
	The next step, before turning to differential atom-interferometric schemes, would be to transfer our results to single atom interferometers with many pulses and to include also the dynamics between the pulses and the corresponding perturbations without electromagnetic interaction.
	Although the effects during the pulse are small in the context of atom interferometers for sufficiently short pulse durations, they might accumulate for multiple sequential pulses and large-momentum-transfer schemes~\cite{Rudolph2020} planned for gravitational-wave and DM detectors~\cite{Graham2013,El-Neaj2020}.
	Our results will help to analyze also much more dominant effects between the pulses and will lead to a comprehensive description of differential atom-interferometric experiments.
	
	\begin{acknowledgments}
		We are grateful to W. P. Schleich for his stimulating input and continuing support.
		We also thank O. Buchmüller, A. Friedrich, J. Rudolph, and C. Ufrecht, as well as the QUANTUS and INTENTAS teams for fruitful and interesting discussions.
		The QUANTUS and INTENTAS projects are supported by the German Space Agency at the German Aerospace Center (Deutsche Raumfahrtagentur im Deutschen Zentrum f\"ur Luft- und Raumfahrt, DLR) with funds provided by the Federal Ministry for Economic Affairs and Climate Action (Bundesministerium f\"ur Wirtschaft und Klimaschutz, BMWK) due to an enactment of the German Bundestag under Grant Nos. 50WM1956 (QUANTUS V), 50WM2250D-2250E (QUANTUS+), as well as 50WM2177-2178 (INTENTAS).
		The projects ``Metrology with interfering Unruh-DeWitt detectors'' (MIUnD) and ``Building composite particles from quantum field theory on dilaton gravity'' (BOnD) are funded by the Carl Zeiss Foundation (Carl-Zeiss-Stiftung).
		The Qu-Gov project in cooperation with ``Bundesdruckerei GmbH'' is supported by the Federal Ministry of Finance (Bundesministerium der Finanzen, BMF).
		F.D.P. is grateful to the financial support program for early career researchers of the Graduate \& Professional Training Center at Ulm University and for its funding of the project ``Long-Baseline-Atominterferometer Gravity and Standard-Model Extensions tests'' (LArGE).
		E.G. thanks the German Research Foundation (Deutsche Forschungsgemeinschaft, DFG) for a Mercator Fellowship within CRC 1227 (DQ-mat).
	\end{acknowledgments}
	
	\section*{Author declarations}
	\subsection*{Conflict of interest}
	
	\noindent The authors have no conflicts to disclose.
	
	\subsection*{Author contributions}
	
	\textbf{Alexander Bott} formal analysis (lead); investigation (lead); methodology (lead); software (lead); visualization (lead); writing - original draft (equal); Writing – review and editing (equal).
	\textbf{Fabio Di Pumpo} conceptualization (equal); investigation (supporting); methodology (supporting); project administration (equal); supervision (equal); validation (equal); visualization (supporting); writing - original draft (equal); Writing – review and editing (equal).
	\textbf{Enno Giese} conceptualization (equal); investigation (supporting); methodology (supporting); project administration (equal); supervision (equal); validation (equal); visualization (supporting); writing - original draft (equal); Writing – review and editing (equal).
	
	\section*{Data availability}
	The data that support the findings of this study are available within the article.
	
	\appendix
	\section{Adiabatic Elimination}
	\label{sec:adel}
	The three-level Hamiltonian for magnetically-induced single-photon transitions from Eq.~\eqref{3levelHam} is of the form 
	\begin{equation}\label{key}
		\hat{H}=\hbar\begingroup
		\begin{pmatrix} \hat{\nu}_{a}+\Delta & \hat{T}^\dagger \\
			\hat{T} & \hat{\nu}_{eg}
		\end{pmatrix}\endgroup,
	\end{equation}
	where the free evolution of the two-level state $\ket{\psi_{eg}}=(\ket{\psi_{e}},\ket{\psi_g})^T$ and the ancilla state $\ket{\psi_a}$ are determined by $\hat{\nu}_{eg}$ and $\hat{\nu}_{a}+\Delta$.
	Transitions between the two systems are described by $\hat{T}$ and all quantities except $\Delta$ may depend on time.
	We now define the projector~\cite{Sanz2016} $\hat{P}$ such that $\ket{\psi_a}=\hat{P}\ket{\psi_{eg}}$.
	\begin{table*}
		\caption{\label{Table:CoeffDetMean}
			Coefficients $\hat{\nu}^{(j)}$ and $\hat{\bar{\nu}}^{(j)}$ of the detuning $\hat{\nu}_\text{H}=\sum_{j=0}^{3}\hat{\nu}^{(j)}t^j$ and mean energy shift $\hat{\bar{\nu}}_\text{H}=\sum_{j=0}^{3}\hat{\bar{\nu}}^{(j)}t^j$ in the Heisenberg picture expanded in powers of time.
			Here, we have assumed that the dark matter field is constant on the timescale of the pulse, \ie{} $\varrho_{\text{DM}}(\hat{z}_{\text{H}},t)\cong\varrho_{\text{DM}}(\hat{z},0)$, with perturbative parameters $\omega_{\varrho}/\Omega$, $k_{\varrho}\hat{p}/(\bar{m}\Omega)$, and $k_{\varrho}g/\Omega^2\ll1$.
			If the unperturbed resonance condition is not met perfectly or velocity selectivity arises, its residual contributes to the effective detuning and is time independent, as expected.
			Similarly, the Doppler effect as well as chirping lead to a detuning that depends linearly on time.
			Besides these known contributions, other perturbations are included that in addition arise from gravity ($\sim g$), chirping ($\sim\alpha$), the mass defect ($\sim\omega_{eg}$), EEP violation ($\sim\beta_S$) and dark matter ($\sim\varrho_{\text{DM}}$).
			The mean energy is solely determined by these perturbations.
		}
		\begin{ruledtabular}
			\begin{tabular}{ccc}
				$t^j$  & Coefficients $\hat{\nu}^{(j)}$ of the detuning $\hat{\nu}_{\text{H}}$                            & Coefficients $\hat{\bar{\nu}}^{(j)}$ of the mean energy $\hat{\bar{\nu}}_{\text{H}}$  \\ 
				\hline
				\rule{0pt}{5ex} $t^0$  & \makecell{$(\omega_{eg}+\hat{\nu}_k+\Delta\omega_{ac})-\omega_\text{L}+\frac{k\alpha\hat{z}}{c}+\frac{\omega_{eg}}{\bar{\omega}}\left(\frac{\bar{m}g\hat{z}}{\hbar}-\frac{\hat{p}^2}{2\bar{m}\hbar}-\frac{\omega_k}{4}\right)$\\$+\bar{\omega}\Delta\beta\varrho_{\text{DM}}(\hat{z},0)+\bar{\omega}\Delta\beta\beta_S\frac{g\hat{z}}{c^2}-\left\{\frac{k\hat{p}}{2\bar{m}},\frac{(g+\alpha)\hat{z}}{c^2}\right\}$}  & $\bar{\omega}\bar{\beta}\varrho_{\text{DM}}(\hat{z},0)+\bar{\omega}\bar{\beta}\beta_S\frac{g\hat{z}}{c^2}-\frac{k\hat{p}}{4\bar{m}}\frac{\omega_{eg}}{\bar{\omega}}-\frac{\omega_k}{2}\frac{(g+\alpha)\hat{z}}{c^2}$  \\
				\rule{0pt}{4ex} $t^1$  & $-k(g+\alpha)\left(1+\frac{\hat{p}^2}{\bar{m}^2c^2}-\frac{g\hat{z}}{c^2}\right)+2\hat{\nu}_k\frac{\alpha}{c}+\left(2\frac{\omega_{eg}}{\bar{\omega}}+\Delta\beta\beta_S\right)\frac{g\hat{p}}{\hbar}$  & $\bar{\beta}\beta_S\frac{g\hat{p}}{\hbar}+\frac{kg}{4}\frac{\omega_{eg}}{\bar{\omega}}+\frac{\omega_k}{2}\left(\frac{\alpha}{c}-\frac{g+\alpha}{c^2}\frac{\hat{p}}{\bar{m}}\right)$\\
				\rule{0pt}{4ex} $t^2$  & $-\frac{3kg\alpha}{2c}-\frac{\omega_{eg}}{\bar{\omega}}\frac{\bar{m}g^2}{\hbar}-\bar{\omega}\Delta\beta\beta_S\frac{g^2}{2c^2}+\hat{\nu}_k\frac{g(g+\alpha)}{c^2}$  & $-\bar{\omega}\bar{\beta}\beta_S\frac{g^2}{2c^2}+\frac{\omega_k}{2}\left(\frac{\alpha^2}{c^2}+\frac{g+\alpha}{c^2}\frac{g}{2}\right)$\\
				\rule{0pt}{4ex} $t^3$  & $-k(g+\alpha)\frac{g^2}{c^2}$ & $0$\\
			\end{tabular}
		\end{ruledtabular}
	\end{table*}
	Note that $\hat{H}$ depends on time, so $\hat{P}$ may also be time dependent.
	Without this projector, we find the Schr\"odinger equations 
	\begin{subequations}
		\begin{align}
			\ii\frac{\partial}{\partial t}\ket{\psi_{a}} &= \left(\hat{\nu}_a+\Delta\right)\ket{\psi_a}+\hat{T}^\dagger\ket{\psi_{eg}} \label{SGancillaState} \\ 
			\ii\frac{\partial}{\partial t}\ket{\psi_{eg}} &= \hat{T}\ket{\psi_a}+\hat{\nu}_{eg}\ket{\psi_{eg}} \label{SGlowerStates}  
		\end{align}
	\end{subequations}
	for the ancilla state and the two lower levels.
	Then, we use $\ket{\psi_a}=\hat{P}\ket{\psi_{eg}}$ to eliminate the ancilla state, resulting in
	\begin{subequations}
		\begin{align}
			\ii\frac{\partial}{\partial t}\ket{\psi_{eg}} &= \hat{P}^{-1}\left[\left(\hat{\nu}_a+\Delta\right)\hat{P}+\hat{T}^\dagger-\ii\frac{\partial}{\partial t}\hat{P}\right]\ket{\psi_{eg}} \label{SGancillaState} \\ 
			\ii\frac{\partial}{\partial t}\ket{\psi_{eg}} &= \Bigl[\hat{T}\hat{P}+\hat{\nu}_{eg}\Bigr]\ket{\psi_{eg}}. \label{SGlowerStates}  
		\end{align}
	\end{subequations}
	Both equations describe the time evolution of the two lower internal states, so that the Hamiltonians as defined by the right-hand sides must be the same.
	Thus, we obtain the Bloch equation for the projector $\hat{P}$ 
	\begin{equation}\label{key}
		\Delta \hat{P}+\hat{T}^\dagger = -\hat{\nu}_a\hat{P}+\hat{P}\hat{\nu}_{eg}+\ii\frac{\partial}{\partial t}\hat{P}+\hat{P}\hat{T}\hat{P}
	\end{equation}
	from equating the right-hand sides, including a time derivative of the projector due to the time dependence of the Hamiltonian.
	The solution of the equation can be found by employing the ansatz $\hat{P} = \sum_{n=0}^{\infty}\hat{P}_n$.
	We search for a perturbative expansion where $||\hat{P}_{n+1}||\ll||\hat{P}_n||$, in order to truncate the series.
	For that, we assume that $\Delta$ is much larger than all other frequency scales of the Hamiltonian, leading to the expansion parameters $||\hat{\nu}_{a}/\Delta||$, $||\hat{\nu}_{eg}/\Delta||$, and $||\hat{T}/\Delta||$.
	Furthermore, we assume that the frequency scales introduced by chirping and the dilaton are also small compared to $\Delta$, \ie{} $||\dot{\hat{\nu}}_{a}/\Delta^2||$, $||\dot{\hat{\nu}}_{eg}/\Delta^2||$, and $|| \dot{\hat{T}}/\Delta^2||$ can even be treated as of higher order than the parameters above.
	Consequently, $||\dot{\hat{P}}_{n}/\Delta||$ is also of higher order.
	
	As outlined already in the main part, we search for the relation $||\hat{P}_{n+1}||\sim||\hat{P}_n/\Delta||$, allowing us to truncate the series.
	If additionally $||\hat{P}_0||\sim | \Delta^{-1}|$ holds, the right-hand side of the Bloch equation becomes smaller than the left-hand side by at least one order in the expansion parameters. As a consequence, we neglect terms on the right-hand side and find the lowest order of the projector series
	\begin{equation}\label{key}
		\hat{P}_0 = -\frac{1}{\Delta}\hat{T}^\dagger,
	\end{equation}
	which fulfills the condition $||\hat{P}_0 ||\sim |\Delta^{-1}|$.
	The next order can be obtained by inserting the series representation of the projector into the Bloch equation and by cancelling the lowest order from both sides to obtain
	\begin{equation}
		\hat{P}_1 = \frac{1}{\Delta^2}\left(\hat{\nu}_a\hat{T}^\dagger-\hat{T}^\dagger\hat{\nu}_{eg}-\ii\frac{\partial}{\partial t}\hat{T}^\dagger\right),
	\end{equation}
	fulfilling the relation $||\hat{P}_{1}||\sim||\hat{P}_0/\Delta||$ as desired.
	By inserting this result into the Bloch equation and by applying the Cauchy product to the double sum, we find the recursion formula for higher orders 
	\begin{equation}\label{key}
		\hat{P}_{n+1} = \frac{1}{\Delta}\left(-\hat{\nu}_a\hat{P}_{n}+\hat{P}_{n}\hat{\nu}_{eg}+\ii\frac{\partial}{\partial t}\hat{P}_{n}+\sum\limits_{k=0}^{n-1}\hat{P}_{k}\hat{T}\hat{P}_{n-k}\right)
	\end{equation}
	for $n\geq 1$.
	Hence, the desired behaviour of $||\hat{P}_{n+1}||\sim||\hat{P}_n/\Delta||$ is achieved by the above choice of $\hat{P}_0$.
	From the Schr\"odinger equation for the two lower states in the form of Eq.~\eqref{SGlowerStates} follows the effective two-level Hamiltonian 
	\begin{equation}
		\label{key}
		\hat{\mathcal{H}}_{\text{rot}} = \hbar\left(\hat{T}\hat{P}+\hat{\nu}_{eg}\right)\cong-\frac{\hbar}{\Delta}\hat{T}\hat{T}^\dagger+\hbar\hat{\nu}_{eg}
	\end{equation}
	to lowest order in $\Delta^{-1}$.
	Including further corrections of higher orders than $\Delta^{-1}$ would require a symmetrization of the Hamiltonian~\cite{Sanz2016}.
	
	\section{Time evolution in Heisenberg picture}
	\label{sec:Heisenberg}
	\renewcommand\dbltextfloatsep{20.0pt}
	The resulting Hamiltonian in the Heisenberg picture
	\begin{equation}
		\label{key}
		\hat{\mathcal{H}}_{\text{H}} = \frac{\hbar}{2}\renewcommand*{\arraystretch}{2}
		\begin{pmatrix}
			2\hat{\bar{\nu}}_{\text{H}}+\hat{\nu}_{\text{H}} & \Omega\\
			\Omega & 2\hat{\bar{\nu}}_{\text{H}}-\hat{\nu}_{\text{H}}
		\end{pmatrix}
	\end{equation}
	after adiabatic elimination includes the detuning $\hat{\nu}_{\text{H}}$ and the mean energy $\hat{\bar{\nu}}_{\text{H}}$ as given in Eqs.~\eqref{eq:DetInHeis} and \eqref{nuHbar}.
	We assume that the unperturbed resonance condition is sufficiently fulfilled such that $\hat{\nu}_\text{H}$ and $\hat{\bar \nu}_\text{H}$ can be treated as perturbative quantities, \ie{} $\Omega\gg ||\hat{\nu}_\text{H}||,||\hat{\bar \nu}_\text{H}||$.
	Hence, the dominant frequency contribution to the Hamiltonian is given by the Rabi frequency $\Omega$.
	We remove this contribution~\cite{Bertoldi2019} by changing into a picture co-rotating with $\Omega$ via the transformation
	\begin{equation}
		\label{key}
		\hat{U}_{\varphi} = \renewcommand*{\arraystretch}{2}
		\begin{pmatrix}
			\cos\varphi_t & -\ii\sin\varphi_t\\
			-\ii\sin\varphi_t & \cos\varphi_t
		\end{pmatrix}
	\end{equation}
	with (half) pulse area $\varphi_t = \Omega t/2$.
	This way, we find the transformed Hamiltonian 	
	\begin{equation}
		\label{key}
		\hat{\mathcal{H}}_{\varphi} = \frac{\hbar}{2}\renewcommand*{\arraystretch}{2}
		\begin{pmatrix}
			2\hat{\bar{\nu}}_{\text{H}}+\hat{\nu}_{\text{H}}\cos2\varphi_t & -\ii\hat{\nu}_{\text{H}}\sin2\varphi_t\\
			\ii\hat{\nu}_{\text{H}}\sin2\varphi_t & 2\hat{\bar{\nu}}_{\text{H}}-\hat{\nu}_{\text{H}}\cos2\varphi_t
		\end{pmatrix}.
	\end{equation}
	For the generation of $\pi$ and $\pi/2$ pulses as mirrors and beam splitters, the typical interaction duration is proportional to $\Omega^{-1}$.
	We observe that the transformed Hamiltonian is perturbative on these time-scales, \ie{} $|| \hat{\mathcal{H}}_{\varphi}||/(\hbar\Omega)\ll1$.
	Thus, we can use a first-order Dyson expansion to determine the time evolution in the co-rotating picture.
	Reverting the transformation $\hat{U}_{\varphi}$ after performing the expansion yields the time evolution in the Heisenberg picture
	\begin{equation}
		\label{key}
		\hat{U}= \renewcommand*{\arraystretch}{2}
		\begin{pmatrix}
			\hat{C}_1\cos\varphi_{t}-\hat{S}\sin\varphi_{t} & 	-\ii\hat{C}_0\sin\varphi_{t}+\ii\hat{S}\cos\varphi_{t}\\
			-\ii\hat{C}_1\sin\varphi_{t}-\ii\hat{S}\cos\varphi_{t} & \hat{C}_0\cos\varphi_{t}+\hat{S}\sin\varphi_{t}
		\end{pmatrix}
	\end{equation}
	with operators
	\begin{subequations}\label{CS}
		\begin{align}
			\hat{C}_{1/0} &= 1-\frac{\ii}{2}\int\limits^{t}_{0}\!\!\!\dd \tau\left[2\hat{\bar{\nu}}_{\text{H}}(\tau)\pm\hat{\nu}_{\text{H}}(\tau)\cos2\varphi_\tau\right]\\
			\hat{S} &= \frac{\ii}{2}\int\limits^{t}_{0}\!\!\!\dd \tau\hat{\nu}_{\text{H}}(\tau)\sin2\varphi_\tau
		\end{align} 
	\end{subequations}
	that include integrals over the time-dependent detunings and trigonometric functions.
	\begin{table*}
		\caption{\label{Table:CoeffTimeEvol}
			Coefficients that contribute to the matrix elements of the time evolution in the Heisenberg picture with (half) pulse area $\varphi_t =\Omega t/2$, calculated through the Dyson series. 
			The coefficients determine the influence of the detuning $\hat{\nu}_{\text{H}}$ on the time-evolution, where the $j$th coefficients are prefactors of the operators $\hat{\nu}^{(j)}$ defined in table~\ref{Table:CoeffDetMean}.
			Here, $\eta_j(t)$ contributes to the diagonal elements of the time-evolution operator, $\xi_j(t)$ to the off-diagonal elements.
			Notably, to zeroth order there is no contribution to the off-diagonal elements so that a constant detuning has no effect on the transition between internal states.
		}
		\begin{ruledtabular}
			\begin{tabular}{ccc}
				$j$  & Coefficients $\eta_j(t)$ for diagonal elements $\bra{n}\hat{U}\ket{n}$                    & Coefficients $\xi_j(t)$ for off-diagonal elements $\bra{n}\hat{U}\ket{m\neq n}$\\ 
				\hline
				0  & $2\sin\varphi_{t}$                                                                        & $0$ \\
				1  & $2\varphi_{\tau}\sin\varphi_{t}$                                                           & $-2\sin\varphi_{t}+2\varphi_{t}\cos\varphi_{t}$\\
				2  & $-4\sin\varphi_{t}+4\varphi_{t}\cos\varphi_{t}+4\varphi_{t}^2\sin\varphi_{t}$                 & $4\varphi_{t}(-\sin\varphi_{t}+\varphi_{t}\cos\varphi_{t})$\\
				3  & $-12\varphi_{t}\sin\varphi_{t}+12\varphi_{t}^2\cos\varphi_{t}+8\varphi_{t}^3\sin\varphi_{t}$ & $2(4\varphi_{t}^2-6)(-\sin\varphi_{t}+\varphi_{t}\cos\varphi_{t})-4\varphi_{t}^2\sin\varphi_{t}$\\
			\end{tabular}
		\end{ruledtabular}
	\end{table*}
	For $t\sim 1/\Omega$, a (partial) integration of the time-dependent DM field $\varrho_{\text{DM}}$ leads to contributions of the order $\mathcal{O}(\bar{\varrho}_0t^2)$.
	We neglect these contributions and evaluate the DM part at the initial time, \ie{} $\varrho_{\text{DM}}(\hat{z}_{\text{H}},\tau)\cong\varrho_{\text{DM}}(\hat{z},0)$, treating the DM field as constant during the interaction. 
	The detuning and mean energy in the Heisenberg picture can then be cast into the form $\hat{\nu}_\text{H}=\sum_{j=0}^{3}\hat{\nu}^{(j)}t^j$ and $\hat{\bar{\nu}}_\text{H}=\sum_{j=0}^{3}\hat{\bar{\nu}}^{(j)}t^j$, respectively, with coefficients $\hat{\nu}^{(j)}$ and $\hat{\bar{\nu}}^{(j)}$ as given in table~\ref{Table:CoeffDetMean}.
	Carrying out the integrations in Eq.~\eqref{CS} leads to the time evolution in the Heisenberg picture with matrix elements $\hat{U}_{n,m} = \bra{n}\hat{U}\ket{m}$, with $m,n \in\{e,g\}$.
	The diagonal elements are given by
	\begin{subequations}
		\label{eq:matrixElU}
		\begin{equation}
			\begin{split}
				\hat{U}_{n, n}=& \cos\varphi_{t}\left[1-\ii\sum\limits_{j=0}^{3}\frac{\hat{\bar{\nu}}^{(j)}t^{j+1}}{j+1}\right]+\frac{\ii\lambda_n}{2\Omega}\sum\limits_{j=0}^{3}\frac{\hat{\nu}^{(j)}\eta_j(t)}{\Omega^{j}}
			\end{split}
		\end{equation}
		and the off-diagonal elements by
		\begin{equation}
			\begin{split}
				\hat{U}_{n,m}=&-\ii\sin\varphi_{t}\left[1-\ii\sum\limits_{j=0}^{3}\frac{\hat{\bar{\nu}}^{(j)}t^{j+1}}{j+1}\right]-\frac{\lambda_n}{2\Omega}\sum\limits_{j=0}^{3}\frac{\hat{\nu}^{(j)}\xi_j(t)}{\Omega^{j}}
			\end{split}
		\end{equation}  
	\end{subequations}
	with $\lambda_{e/g}= \mp 1$.
	In these matrix elements, the coefficients $\eta_j(t)$ and $\xi_j(t)$ result from the integration of trigonometric terms combined with the time dependence of the detuning and the mean energy and are given in table~\ref{Table:CoeffTimeEvol}. 
	Notably, the zeroth-order coefficient for the off-diagonal elements vanishes. 
	A constant detuning therefore does not influence the off-diagonal elements of the time-evolution, which describe the transition between internal atomic states.
	At $\varphi_{t} = 0$ we furthermore find $\hat{U} = \mathbbm{1}$ as expected.
	The pulse areas of interest are those of mirror pulses, where $\varphi_{t} = \pi/2$, and beam-splitter pulses , where $\varphi_{t} = \pi/4$, in which the form of $\hat{U}_{n,m}$ given in Eq.~\eqref{eq:matrixElU} is valid.
	\begin{widetext}
		\section{Phase from modified wave vector}
		\label{sec:phase}
		In this Appendix, we present the phase difference of a Gaussian wave packet for a mirror pulse after diffraction that originates from the wave-vector modification together with imperfect chirping ($\alpha\neq-g$), generalizing Eq.~\eqref{eq:PhaseWaveVEc} from the main part.
		With momenta for ground and excited state separated by $\hbar k$ after diffraction, we find
		\begin{align}
			\begin{split}
				\phi_{\text{WV}}=&-\frac{k\alpha t}{c}\left(\bar{z}-\frac{pt}{\bar{m}}-\frac{v_rt}{2}- \frac{gt^2}{2}\right)-\frac{k\alpha}         
				{\Omega^2}\left(\frac{2p}{\bar{m}c}+\frac{v_\text{r}}{c}+\frac{gt}{c}\right) \\
				&+\frac{k\left(g+\alpha\right)}{2c^2}\left[\left( \frac{p_{e}-p-\hbar k-\bar m gt}{\hbar \sigma_e^2}\right)^2-\frac{1}{\sigma_e^2}+\left( \frac{p_{g}-p-\bar m gt}{\hbar \sigma_g^2}\right)^2-\frac{1}{\sigma_g^2}-\left(\bar{z}-\frac{pt}{\bar{m}}-\frac{v_rt}{2}-\frac{gt^2}{2}\right)^2-\bar{z}^2-\frac{\Delta z^2}{2}\right]\\
				&+\frac{\Delta z\left(g+\alpha\right)}{2c^2}
				\omega_k t+\frac{k(g+\alpha)}{\Omega^2}\Bigg[2\left(\frac{p}{\bar{m}c}+\frac{v_r}{2c}\right)^2+\frac{pgt}{\bar{m}c^2}+\frac{v_rgt}{2c^2} +2\frac{g\bar{z}}{c^2}+\frac{\pi^2 - 12}{2\pi^2}\left(\frac{gt}{c}\right)^2\Bigg].
			\end{split}
		\end{align}
		All but the first line cancel in the case of perfect chirping.
		Moreover, we find that only if there is a chirping mismatch, the finite width of the wave packets is relevant.
	\end{widetext}
	\bibliography{Literatur}
	
\end{document}